\documentclass[%
 reprint,
showpacs,preprintnumbers,
 amsmath,amssymb,
 aps,
]{revtex4-1}

\usepackage{graphicx}% Include figure files
\usepackage{dcolumn}% Align table columns on decimal point
\usepackage{bm}% bold math
\begin{document}

%\preprint{APS/123-QED}

\title{Quantum phases of a one-dimensional dipolar Fermi gas}% Force line breaks with \\
\author{Hamid Mosadeq$^{1,2}$}
\author{Reza Asgari$^2$}
 \email{asgari@ipm.ir}
\affiliation{$^1$ Faculty of Science, Shahrekord University, Shahrekord 88186-34141, Iran }
\affiliation{$^2$ School of Physics, Institute for Research in Fundamental Sciences (IPM), Tehran 19395-5531, Iran}

\date{\today}% It is always \today, today,

\begin{abstract}
We quantitatively obtain the quantum ground-state phases of a Fermi system with on-site and dipole-dipole interactions in one-dimensional lattice chains within the density matrix renormalization group. We show, at a given spin polarization, the existence of six phases in the phase diagram and find that the phases are highly dependent on the spin degree of freedom. These phases can be constructed using available experimental techniques.
\end{abstract}
\maketitle

\section{introduction}\label{sect:intro}

\begin{figure}[t]
\includegraphics[width=0.9\linewidth]{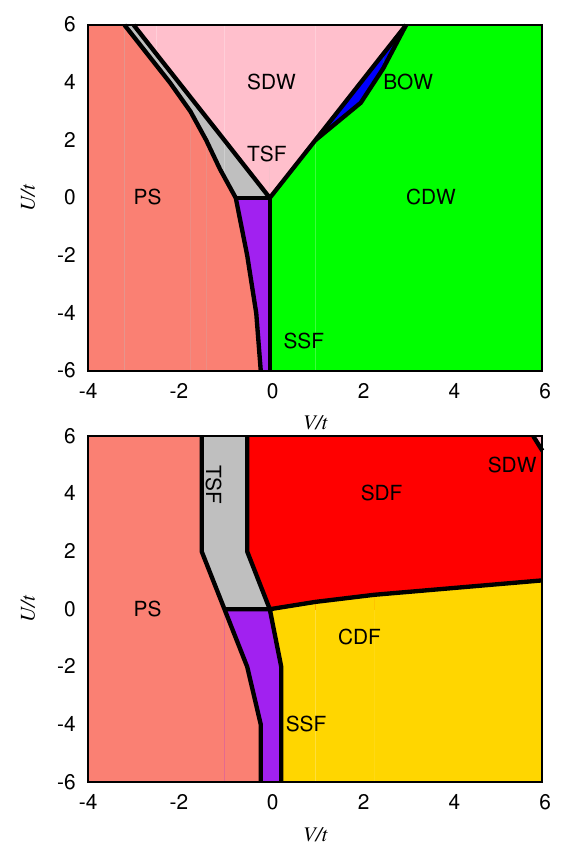}
\caption{(Color online) DMRG phase diagrams of the fermion atoms with dipolar interactions in the unpolarized 1D lattice chain. At half filling, ( top panel) a quantum TSF phase occurs mainly for all $0<U/t<3$ and $-0.5<V/t<0$ whereas the SSF phase mostly takes place for $-0.5<V/t<0$ and $U/t<0$. The BOW is located in a narrow strip between the spin and charge density wave phases. The PS phase occurs for the region where $V/t<-0.75$ and the boundary disperses as a function of $U$. For the attractive dipolar interaction values, an inhomogeneous cluster type phase occurs. For the case at quarter filling ( bottom panel) rich quantum phases including a large area of the charge density fluctuation (CDF) for $0<V/t$ and $U<V/6$, the SSF phase mainly for $U<0$ and $-1<V/t<0.25$, spin density fluctuation phase (SDF) for the positive $U$ and $V$, and a large area of PS occur. In addition, the TSF phase is located around small negative $V$ and $U>0$ next to the spin density fluctuation phase. A SDW phase occurs for large $U/t>6$ and $V/t>6$ values. The mesh in the horizontal axis is $\Delta V=0.05t$ and the different phases are separated by lines where their thickness is much larger than numerical errors.
}
\label{fig1}
\end{figure}

Cold dipolar atom gas systems have
attracted a lot of attention due to the novel anisotropic and
long-range character of the dipole-dipole interactions~\cite{Baranov}.
For high enough densities, the atomic de Broglie wavelength
becomes larger than the typical inter-particle distance and thus
quantum statistics governs the many-body dynamics of cold atom
systems. Moreover, for strong fermion-fermion interactions, when
the average interaction energy becomes larger than the
corresponding kinetic energy, one can expect drastic changes in 
the properties of the system. Strong correlations are at the
center of activity of various scientific disciplines such as optical, condensed matter physics, chemistry and quantum science
ranging from high-temperature superconductivity~\cite{Leggett1},
superfluidity (SF)~\cite{Leggett}, metal-insulator
transition~\cite{Imada}, Fulde-Ferrel-Larkin-Ovchinkov
(FFLO)~\cite{fflo}, orbital ordering and other structural
phase transitions~\cite{Cowley}.

One of the current challenges of condensed matter physics is to
understand the distinctive exotic paired states and quantum phases that are realized
when particles have different on-site and long-range
interactions. It has been predicted that in Bose lattice systems,
the presence of finite interactions gives rise to novel quantum
phases in two-dimensional~\cite{Sansone} and one-dimensional
(1D)~\cite{Kumar, 1D-gas} systems. A quantum phase
diagram of fermionic dipolar gases in a planar array of one-dimensional tubes has been studied~\cite{huang} and the elementary excitations and
the Luttinger components for various correlation functions were found. Unconventional SF in two coupled fermionic chains has been proposed~\cite{Uchino} in which an admixture of spin singlet and triplet SF pairings occurres with purely repulsive interactions. The recent experimental investigations~\cite{Exp, Exp0, Exp1} in creating degenerate cold polar molecules, relying on the dipole-dipole interaction and using the many internal degrees of freedom in molecules to engineer effective spin-spin interactions offer promising orientations for exploring novel and strong correlated many-body physics. More importantly, dipolar interactions can enrich considerably the physics of quantum gases with internal degrees of freedom. The experimental observation of dipolar systems started with reporting the realization of a chromium Bose-Einstein condensate with strong dipolar interactions. By using a Feshbach resonance, Lahaye et al.~\cite{Exp} reduced the usual isotropic contact interaction, such that the anisotropic magnetic dipole-dipole interaction between $^{52}$Cr atoms becomes comparable in strength. Afterwards, the creation of an ultracold dense gas of potassium rubidium polar molecules was reported~\cite{Exp0} and the authors coherently transferred extremely weakly bound potassium rubidium molecules to the vibrational ground state of either the triplet or the singlet electronic ground molecular potential. The range of the dipolar-dipolar interactions can be much larger than the typical optical lattice spacing for systems in which molecules with permanent electric or atomic magnetic dipolar moments have been used~\cite{Exp2}.

In this paper, we employ the density matrix
renormalization group (DMRG)~\cite{code}, which is one of the sophisticated
methods for investigating 1D many-body systems, and finite size scaling to study the phase
diagram of 1D dipolar Fermi systems. Numerical
simulations based on DMRG have been used to investigate the quantum
phases of a 1D Bose lattice~\cite{Torre} but an accurate phase diagram for a 1D dipolar Fermi system is still missing. We find that paired states
near the vanishing on-site energy of a quarter filling
state (one fermion per two sites, $n=1/2$) are significantly
different from those paired states of a half filling state (one
fermion per site, $n=1$). The resulting phase diagram shows the existence of six phases, illustrated in
Fig.~\ref{fig1} for the unpolarized case, which have rich exotic phases of 1D dipolar Fermi gas. The weak coupling phase diagram incorporates spin-density wave (SDW), charge-density wave (CDW), and singlet and triplet superfluidity phases, (SSF or TSF, respectively). In the strong coupling regime, bond order wave (BOW) and phase separation (PS) phases are obtained. Below, we explain the whole states by computing at their
ordered parameters and Luttinger parameters and exploring phase diagrams for polarized phase diagram within DMRG. Notice that the experiments on polar atoms or molecules fall outside the range of validity of the Hubbard model and it is necessary to consider a long-range interaction as a dipolar-dipolar interaction.  We also examine the phase diagrams at finite spin polarization and find that phases are sensitive to the spin degree of freedom, $\xi=(N_{\uparrow}-N_{\downarrow})/N$ where $N=N_{\uparrow}+N_{\downarrow}$ and $N_{\sigma}$ is the number of fermions with spin $\sigma$.

The paper is organized as follows. We introduce the model Hamiltonian and some physical correlation functions in Sec.
\ref{sec:theory}. The numerical
results and discussions regarding phase diagrams, the charge and spin gaps and the density profiles, are reported and discussions regarding different phases at given spin degree of freedom are provided in Sec.~III. A brief summary of results is given in Sec.~\ref{sec:summary}.

\section{Theory and Model}\label{sec:theory}

The fermionic particles
interact with other fermionic particles with an on-site
Coulomb repulsion when two fermions occupy the same orbital. The
Hamiltonian of such interacting ultracold systems on a lattice is
given by the Hubbard model~\cite{Hubbard} which serves as one of
the most prominent models for a solid.
The extended Hubbard model~\cite{EHM, Nakamura} whose particle interaction is modeled by an on-site interaction with a constant interaction potential, describing the low-energy physics of
the interacting dipolar spin-$1/2$ fermionic in a 1D lattice is given by
\begin{eqnarray}
&H& = -t \sum_{i,\sigma} ( c^{\dag}_{i\sigma} c_{i+1,\sigma} + h.c.) + U \sum_i n_{i\uparrow}n_{i\downarrow}\nonumber\\
&+& \sum_{i,r=a}^{(N-1)a}V_i \frac{n_{i}n_{i+r}}{r^3}
\end{eqnarray}
where $c^{\dag}_{i \sigma}$ stands for the creation operator with a spin $\sigma$ at site $i$, $n_{i \sigma}=c^{\dag}_{i \sigma}c_{i \sigma}$ is the density operator and $n_{i}=n_{i \uparrow}+n_{i \downarrow}$. Here $N$ is the cut of the dipolar interaction and we consider $N=7$ such that the physical quantities remain unchanged by increasing $N>7$ and $a$ is a lattice constant (with a typical chemical interaction distance of 1 nm). $t$ represents the transfer energy between the nearest-neighbor sites and $V_i=V$ is the strength of
the dipole-dipole interaction that can change from a positive value to a negative one depending on the direction of the interacting dipolar. It is worth mentioning that a purely attractive interaction can be achieved in the meta-stable state~\cite{trefzger}. The dipole-dipole interaction, in general, has two important features; namely anisotropic and long-ranged, i.e. it decays as $1/r^3$ at large distances. Therefore, it is natural to expect intriguing properties in dipolar gas systems~\cite{asgari}.

We accurately investigate the ground-states and phase diagrams of the system described by Eq.~(1) at half and quarter fillings in terms of the positive and negative values of $U$ and $V$. The variation of $V$ corresponds to a change in the polarization direction with respect to the lattice orientation. Meanwhile, polar molecules are easily manipulated by external electric fields and thus their dipole moments can be tuned.

DMRG is an algorithm for optimizing a variational wavefunction with the structure of a matrix-product state and it consists of a systematic truncation of the system Hilbert space, keeping a small number of important states in a series of subsystems of increasing size to construct wave functions of the full system~\cite{code}. In DMRG the states retained to construct a renormalization group transformation are the most probable eigenstates of a reduced density matrix instead of the lowest energy states kept in a standard numerical renormalization group calculation.

The DMRG with open-end boundary conditions is employed to obtain the ground-state and low-lying excited-states energies and expectation values of order parameters in the thermodynamic limit, $L\rightarrow\infty$. The reason that we use open-end boundary conditions in the system is to reduce the truncation error which is much smaller than with periodic boundary conditions. In order to keep the boundary
effects small, we add additional terms on the boundaries to the Hamiltonian~\cite{boundary} $V n(n_1+n_L)$ so that a particle on the boundary on average has the same potential energy as the rest of the system. In our numerical calculations, we study chains with up to $160$ sites and increase the number of density matrix eigenstates up to $500$ with a sweep number $10$ in order to have the truncation error less than $10^{-9}$ in the fully polarized state and $10^{-6}$ in other states. Therefore, the finite-size scaling analysis based on the $L$ dependence of quantities is needed and we thus perform finite-size scaling for all quantities.

To determine the phase diagrams, several physical expectation values are calculated. We first obtain the charge and spin gaps as follows
\begin{eqnarray}
\Delta_c&=&[E(N_{\uparrow}+1,N_{\downarrow}+1,S_z)+E(N_{\uparrow}-1,N_{\downarrow}-1,S_z)\nonumber\\
&-&2E(N_{\uparrow},N_{\downarrow},S_z)]/2\\
\Delta_s&=&\left[E(N_{\uparrow}+1,N_{\downarrow}-1,S_z+1)-E(N_{\uparrow},N_{\downarrow},S_z)\right]
\end{eqnarray}
where $E(N_{\uparrow},N_{\downarrow},S_z)$ is the
ground-state energy for a given number of atoms with spin-up
(spin-down) $N_{\uparrow}$, ($N_{\downarrow}$) and total spin in
the $z$ direction, $S_z$.

The single-particle spectral function is of particular interest in relation to photoemission results.
We calculate the charge-charge and spin-spin correlation functions
\begin{eqnarray}
S_{(\pm)}(q)=\frac{1}{L}\sum_{jl}
&&e^{iq(j-l)}[<(n_{j\uparrow}\pm n_{j\downarrow})(n_{l\uparrow}\pm n_{l\downarrow})>\nonumber\\
&&-<(n_{j\uparrow}\pm n_{j\downarrow})><(n_{l\uparrow}\pm n_{l\downarrow})>]
\end{eqnarray}
with
$q=2\pi/L$. Notice that the charge-charge and spin-spin correlation functions can be measured using Bragg scattering experiments which provide a clear signature of the phases. Following the Luttinger-liquid (LL)
theory~\cite{Giamarchi, Clay}, the long-wavelength behavior of the
$S_{(\pm)}(q)$ is governed by the LL charge and spin exponents;
$K_{\rho(\sigma)}=\lim_{q\rightarrow 0}\pi S_{\pm}(q)/q$.

A careful extrapolation of the charge-charge and spin-spin correlation functions at a large wavelength limit are necessary to evaluate the correct value of $K_{\rho(\sigma)}$, respectively in the $L\rightarrow \infty$. We analyze various lengths of the lattice size and perform a finite size scaling analysis stemming from the length dependence of the correlation functions. Since the limit $q\rightarrow 0$ limit is very difficult to attain strictly in numerical calculations of finite systems, the values of $K_{\rho}$ and $K_{\sigma}$ calculated from the $q\rightarrow 0$ limit of $S_{\pm}(q)$ are in general slightly larger than their true values~\cite{Clay}. In order to overcome those difficulties, the number of the density matrix eigenstates might be increased and in practice, it is computationally time consuming. Therefore, we use the Tomonaga-Luttinger mode, to calculate the values of $K_{\rho}$ and $K_{\sigma}$. In the Tomonaga-Luttinger
phase, $K_{\rho}$ and $K_{\sigma}$ are given by~\cite{schulz1}
\begin{eqnarray}
K_{\rho}&=&\frac{\pi}{2}n^2 \kappa v_c\nonumber\\
K_{\sigma}&=&\frac{\pi}{2}n^2 \chi v_s
\end{eqnarray}
where $\kappa$ and $\chi$ are the charge compressibility and spin susceptibility and $v_c$, $v_s$ are the charge and spin velocities, respectively. To obtain $K_{\rho}$, we calculate the charge and spin compressibilities from the charge and spin gaps, respectively,
\begin{eqnarray}
\kappa&=&\frac{2}{n^2 L \Delta_c}\nonumber\\
\chi&=&\frac{4}{n^2 L \Delta_s}
\end{eqnarray}
where the velocities are given by
\begin{eqnarray}
v_c&=&[E_1(N_{\uparrow},N_{\downarrow},0)-E_0(N_{\uparrow},N_{\downarrow},0)]/(2\pi/L)\nonumber\\
v_s&=&[E_1(N_{\uparrow}+1,N_{\downarrow}-1,1)-E_0(N_{\uparrow},N_{\downarrow},0)]/(2\pi/L)\nonumber\\
\end{eqnarray}
in which $E_1(N_{\uparrow},N_{\downarrow},S_z)$ is the lowest excitation energy with total spin $S_z$ for finite system size $L$.

\section{Numerical Results and Discussions}\label{sec:discussion}

In this section, we present our main numerical results based on the theory presented in the previous section. Our aim is to explore the phase diagrams of the system in the half and quarter filling cases. We also consider different spin degrees of freedom, $\xi$, and show that the phase diagrams are sensitive to the value of $\xi$.

\begin{figure}
\includegraphics[width=0.9\linewidth]{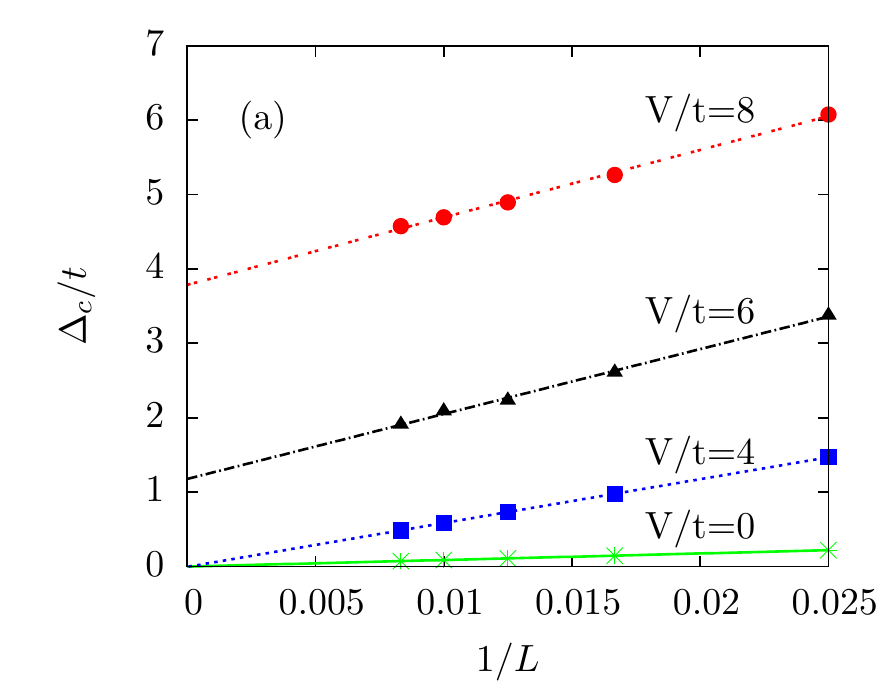}
\includegraphics[width=0.9\linewidth]{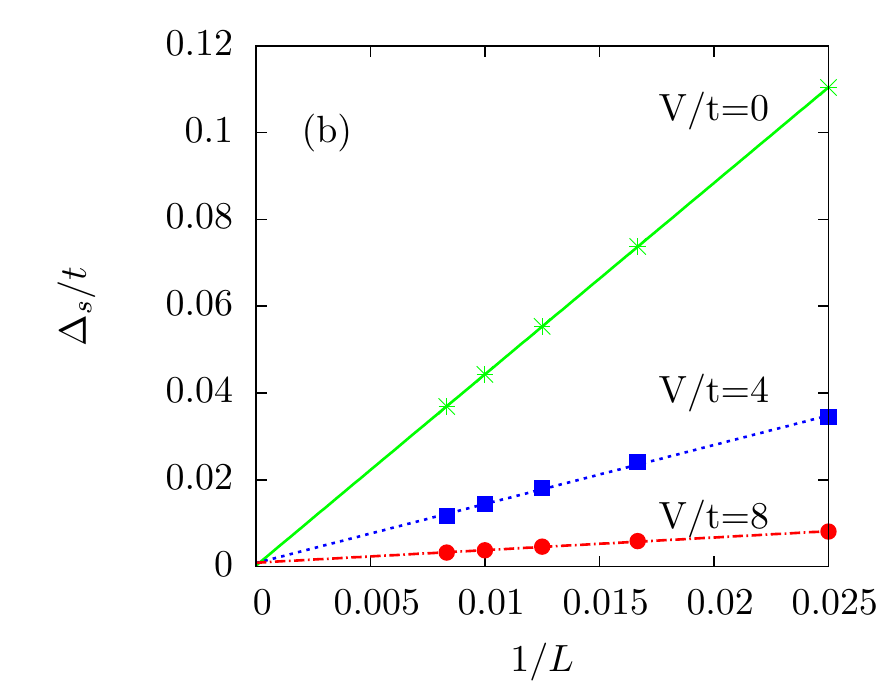}
\caption{(Color online) Finite-size scaling analysis for (a) charge gap  and (b) spin gap scaled by $t$ at $\xi=0$ for $U=V$ and different values of $V$ (in units of $t$) for the case of the quarter filling state, $n=1/2$. The lines indicate the polynomial fits to the numerical results.
}\label{fig2}
\end{figure}

\begin{figure}
\includegraphics[width=0.9\linewidth]{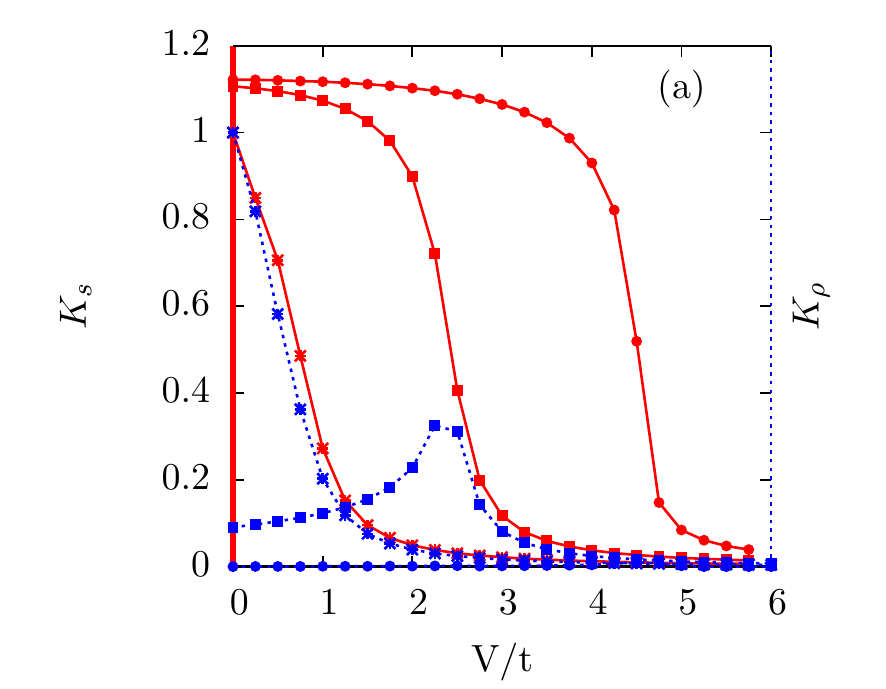}
\includegraphics[width=0.9\linewidth]{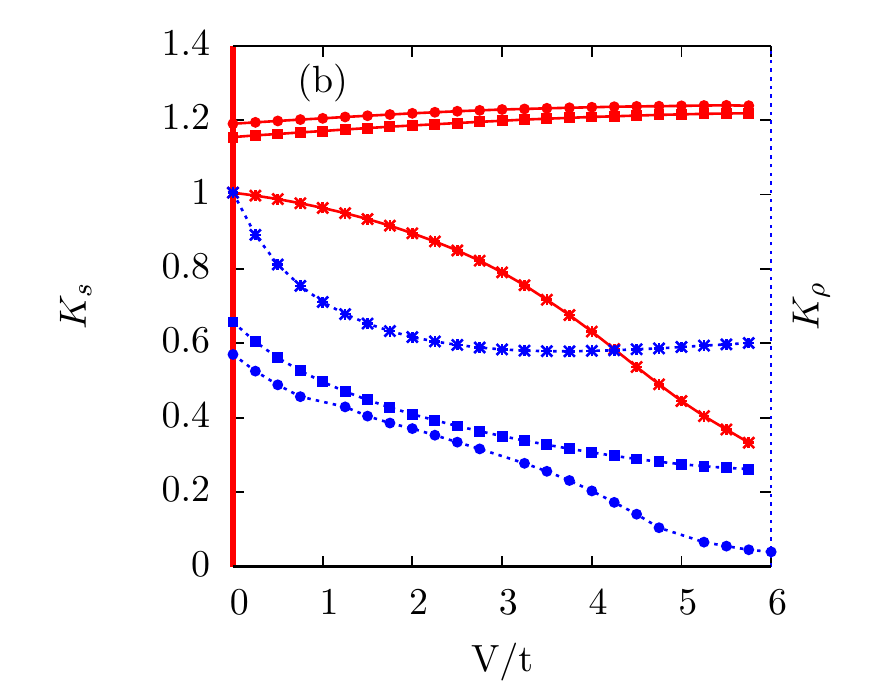}
\caption{(Color online) LL parameters, $K_s$ (red lines) and $K_\rho$ (blue lines), as a function of $V$ (in units of $t$) for different values of the on-site energy for the case of the half filling (a) and the quarter filling states (b) for $\xi=0$ and in the thermodynamic limit. The results for $U/t=0$, $4$ and $8$ are shown by star, squared and circled lines, respectively.}\label{fig3}
\end{figure}

\begin{figure}
\includegraphics[width=0.9\linewidth]{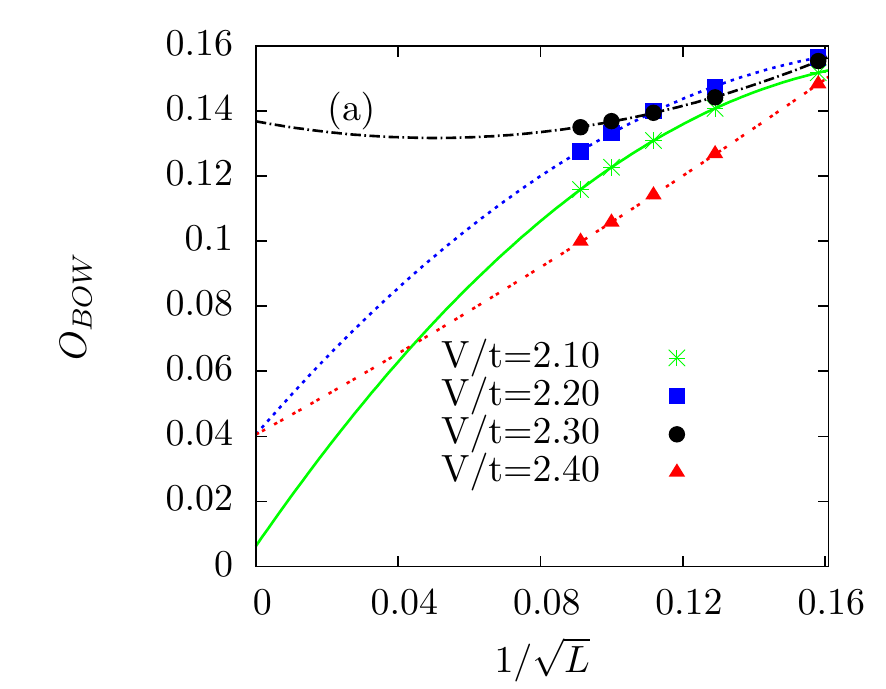}
\includegraphics[width=0.9\linewidth]{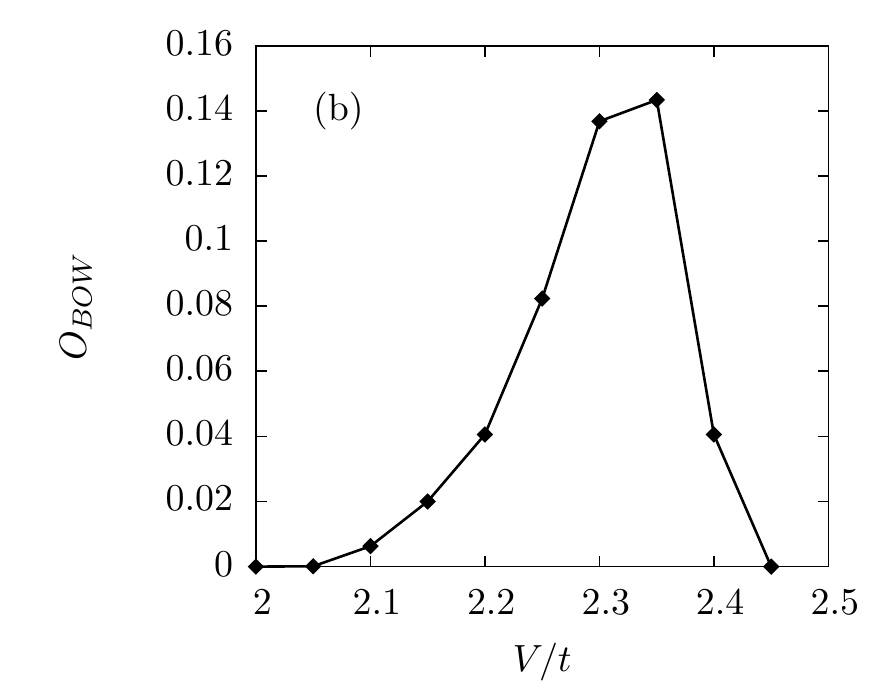}
\caption{(Color online) (a) $O_{BOW}$ quantity as a function of $1/\sqrt{L}$ for different values of the $V$ at $\xi=0$. The lines indicate polynomial fits to the data.(b) $O_{BOW}$ quantity as a function of $V$ (in units of $t$) for the case of half filling at $U=4$.
}\label{fig4}
\end{figure}

\subsection{Unpolarized 1D lattice chain, $\xi=0$}

Having calculated the ground-state energy for a given number of atoms, we can now compute the charge and spin gaps. A careful extrapolation of these quantities is necessary to extract the correct values in the thermodynamic limit, namely $L\rightarrow \infty$. We study various lengths of chains and perform a finite-size scaling analysis based on the $L$ dependence of the physical quantities. Figure~\ref{fig2} shows the finite-size scaling analyses for (a) the charge gap and (b) spin gap at $U=V$ for different values of $V$ for the case with the quarter filling factor $n=1/2$. The value of the gaps depends on the $V$ and $U$ strengths. The charge gap and spin gap vanish for $V/t<4$ , however the charge gap opens for $V/t>4$ in the thermodynamic limit. A finite value of the charge gap or the spin gap indicates different phases and we will discuss those phases in detail later. As shown in Fig.~\ref{fig2}, both of the gaps quantities are a monotonous function of $L^ {-1} $ for all parameters and can be fitted to $\Delta_\delta(L)=a_\delta^{(1)} L^{-1}+a_\delta^{(2)} L^ {-2} $ where $\delta=\rho$ or $\sigma$ denotes the charge or spin channel. We also find that the charge or spin velocity behaves as a line function of $L^{-1}$ and can be fitted nicely to a function $v_\delta(L)=v_\delta^{(1)}/2\pi+v_\delta^{(2)} L^{-1}$.

As we mentioned in the previous section, knowing the Luttinger parameters is vital to understand different phases. $K_\sigma$ vanishes in the spin gapped phase, however $K_\sigma=1$
everywhere else, in the thermodynamic limit. In the weak coupling regime, the metallic phase is a Luttinger liquid~\cite{penc} and it is still quite hypothetical to assume that this is the case for the model Hamiltonian given by Eq.~(1). This assumption is verified by calculating the identity $2 K_{\rho}/(\pi v_{c})=n^2\kappa$ in which $K_{\rho}$ is obtained within the charge-charge correlation function at $q\rightarrow 0$ with a $1$\% error. Moreover, the LL theory has been used for bosonic gases with repulsive power-law interactions~\cite{zoller}. Accordingly, the assumption is applicable in our system in the $(V,U)$ space and we are thus able to explore metallic phases within LL theory.

An exciting phase in 1D systems is a spin density fluctuation (SDF) or Luther-Emery phase, a statistical fluctuation of the spin density where the spin and charge gaps vanish. In order
to determine this phase, we compute the LL exponents~\cite{schulz} where the $K_{\sigma}\ge n^2$ whereas the $K_{\rho}\le n^2$. At the quarter filling, a SDF having the $K_{\rho}\le1/2$ emerges in a wide region of the parameter space and furthermore a small area of spin density wave takes place for large $V$ and $U$ values.

Meanwhile, there is a charge density fluctuation (CDF) phase in which the density correlation function decays
slower than the spin correlation function and thus the density fluctuations are dominant. In this phase, the spin gap is finite and both $K_{\sigma}$ and $K_{\rho}$ are smaller than unity.
When $U\le0$ increases the spin gap opens whereas the charge gap remains zero for certain
values of $V$ at quarter filling. In other words, the ground state of the model is
the CDF metal for a weak interacting fermion regime.

After calculating the charge and spin gaps and corresponding $\Delta_\delta(L)$ together with $v_\delta(L)$, the Luttinger parameters can be evaluated by using Eq.~(5). The charge compressibility from the charge gap is $\kappa=\lim_{L\rightarrow \infty}\kappa(L)=4/n^2 a_{\rho}^{(1)}$ and similarly the spin susceptibility is $\chi=\lim_{L\rightarrow \infty}\chi(L)=4/n^2 a_{\sigma}^{(1)}$. Furthermore, the velocity in the thermodynamic limit can be calculated by using $v_\delta=\lim_{L\rightarrow \infty} v_\delta(L)=v_\delta^{(1)}/2\pi$ and thus we are able to evaluate the Luttinger parameters by using $K_{\delta}=v_{\delta}^{(1)}/a_{\delta}^{(1)}$. Figure ~\ref{fig3} shows $K_{\sigma}$
and $K_\rho$ as a function of $V$ for different values of the on-site energy and for both cases of $n=1$ and $1/2$. Those aforementioned phases can be understood by analyzing $K_{\sigma}$ and $K_\rho$. It should be noted that $K_{\sigma}=K_\rho=1$ for a noninteracting system and the charge parameter, $K_\rho$ increases with increasing $V$ for the case where $U=0$ due to the effective attraction on the same orbital~\cite{penc}. Remarkably, $K_\rho$ tends to zero for the insulating phase such as the CDW as shown in Fig.~\ref{fig3}(a) and it is typically quite difficult to obtain $K_\rho=0$ in an insulating phase near the phase boundary. Having known that $K_{\rho}\neq 0$ on the BOW-CDW boundary (a continuous phase transition) and vanishes elsewhere, we can expect a peak in $K_{\rho}$ as a function of $V$. As shown in Fig.~\ref{fig3}(a), $K_\rho$ decreases from 1 at $U=0$ and shows a peak at $(V,U)=(2.3t,4t)$. Moreover,  $K_\rho$ becomes almost zero for $U>6t$ (the first-order phase transition) in agreement with the results shown in Fig.~1a. Therefore, this behavior explains that the transition is first-order in the $U>6t$ case.

Another state is the PS and it occurs when the
ground-state is inhomogeneous. PS means the possibility of the
system to spontaneously undergo a macroscopic segregation into two
phases with different hole concentrations. In the attractive interaction, a pair of fermion atoms on the same site
can not be broken because, it is costly from an energy point of view. The unpaired fermion particles, on the
other hand, can move in the region
located between pairs, but they can not be on the neighboring sites.
In this phase, the charge gap is finite and $K_{\rho}<n^2$. The simplest way to get quantitative insight into the
instability region is to calculate the compressibility. It turns
out~\cite{Sogo} that the compressibility of a homogenous Fermi gas
becomes negative signaling the instability of the gas that leads to
a collapse. We calculate the compressibility
and thus its divergence illustrates the position of the PS
transition. Our accurate calculations, see Fig.~1, show that a wide regime of the PS phase takes place at both the quarter and half filling phases with an attractive interaction potential.

The competition between the on-site and
dipolar energies gives rise to the stable phase. Recently, by studying the extended Hubbard model ground state
broken symmetries using level crossings in excitation
spectra obtained by exact diagonalization, Nakamura~\cite{Nakamura}
has argued for the existence of a novel bond order wave
(BOW) phase for small to intermediate values of $U$ and $V$
in a narrow strip between the CDW and the SDW phases.
The BOW phase is characterized by alternating strengths
of the expectation value of the kinetic energy operator on
the bonds. It is predicted to be a state where the discrete
symmetry is broken and should hence exhibit
true long-range order.

At half filling, near the bond order
wave~\cite{Nakamura} and charge density wave, which are insulating
phases, the spin gap is suppressed and
moreover the system is a Mott insulator phase with a $2k_{\rm F}$
spin density wave for $V\leq U/2$. Therefore, we introduce local
order parameters for these two gapful phases as $O_{CDW}=<\sum_j
(-1)^j n_j>$ and $O_{BOW}=<\sum_{j \sigma} (-1)^j (
c^{\dag}_{j\sigma} c_{j+1,\sigma} + h.c.)>$. The bond order corresponds to a charge density wave where the density is located on the bonds rather than on the sites as in the CDW. For the finite value of
$O_{CDW}$ or $O_{BOW}$, a long-range order of the charge density wave or bond order wave state
appears. Notice that both charge and spin gaps are
zero in a region near $U=V=0$. In this region, the system is a
gapless Luttinger liquid~\cite{Giamarchi}. In the bond order wave phase with a Mott type insulating gap, both charge and spin gaps are finite.

The BOW order parameter is well extrapolated~\cite{white} as a function of $L^{-K_{\rho}}$. For example, at $U/t=4$ and by sweeping the positive $V$, the SDW phase occurs with zero $K_\rho$ and afterwards we expect $K_{\rho}\simeq 1/2$
in the BOW phase near to the boundary of the SDW phase and by increasing $V$, the CDW phase takes place where the $K_{\rho}$ becomes zero as shown in Fig. 3(a). Similar results have been reported in Ref.~[\onlinecite{white}]. Therefore, since $K_{\rho}\simeq 1/2$, the BOW order parameter is scaled better by $1/\sqrt{L}$ than by $1/L$. The $O_{BOW}$ correlation is shown as a function of $1/\sqrt{L}$ in Fig.~\ref{fig4}(a) at $n=1$ and $U/t=4$. We use polynomial fits to evaluate the quantity in the thermodynamic limit. Afterwards, the $O_{BOW}$ correlation in the $L\rightarrow \infty$ as a function of $V$ is obtained and the results are shown in Fig.~\ref{fig4}(b). The BOW is located in a narrow strip between the spin and charge density wave phases. For a given $2.5<U/t<6$, we find that there is a domain around the $V/t=2$ for which the $O_{BOW}$ of the system is finite in the $L\rightarrow \infty$.
Furthermore, as $V$ increases, the charge fluctuations enhance and thus a
transition from a spin density wave to a bond order wave occurs. The boundary is determined
where the spin gap begins to develop. The occurrence of the bond order wave can be understood as the result of increasing frustration in the spin degree of freedom.

Our numerical results of the phase diagram at half filling and for $V>0$, {\it apart from the bond order  wave phase}, are in good agreement with those results
obtained by a bosonization theory in Ref.~[\onlinecite{Silva}]. Moreover, the results show that there is a discrepancy between the quantum phases at half and quarter fillings due to the difference between the on-site and dipolar energies for different fillings. Noticeably, the phase diagrams shown in Fig.~1 are different from those results obtained by the extended Hubbard model~\cite{penc, tsuchiizu} due to the impact of the dipole-dipole interaction. Essentially the phase diagram of the 1D dipolar system at $n=1/2$ differers with that obtained within the extended Hubbard model. For instance, an insulator phase with a charge gap has been predicted within the extended Hubbard model at quarter filling for the regime in which $U/t>6$ and $V/t>4$ however it does not exist in a 1D dipolar system as shown in the bottom panel of Fig.~1.

\subsection{Polarized 1D lattice chain, $\xi \neq 0$}

The FFLO phase~\cite{fflo} has recently
attracted a lot of interest from both experimental and
theoretical groups~\cite{all} for spin polarized systems. To obtain the FFLO phase, the
pairing operator
$\hat{\Delta}_l=\hat{c}_{l\downarrow}\hat{c}_{l\uparrow}$ is no
longer useful since a long-range order is forbidden in 1D however,
the correlation functions do not decay exponentially
but as power laws, which is very slowly.
The correlation functions of the pairing operator
\begin{equation}
C_{l\acute{l}}=\langle\hat{\Delta}_{l}^{\dag}\hat{\Delta}_{\acute{l}}^{}\rangle
\end{equation}
for different values of the spin polarization can be evaluated and
our numerical results for the polarized system, $\xi\neq 0$, show that the pair correlation function
decays with a power law $|l-\acute{l}|^{-1/K_{\rho}}$ at large
distances. As the value of $\xi$ increases,
the power law of the correlation function transforms to
an oscillation function at large distances. For $V<0$, the form of the function differs with respect to
the $V>0$. Moreover the pair correlation
function increases by decreasing the value of $U$.

\begin{figure}
\includegraphics[width=1\linewidth]{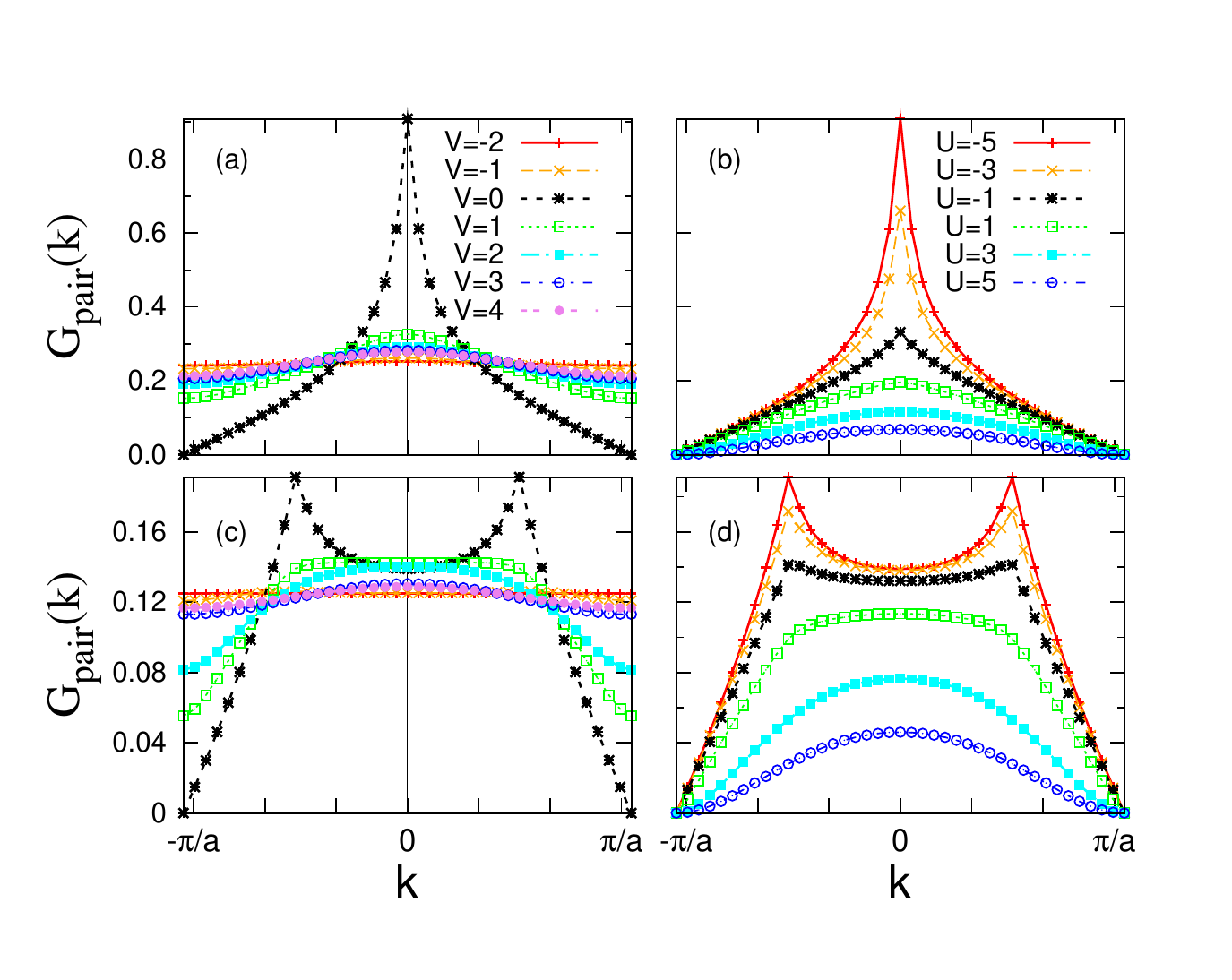}
\caption{(Color online) DMRG pair momentum distribution as a function of $k$ for a system with $L=40a$ at (a) $U/t=-5$ for the potential energy $-2\leq V/t\leq4$ and at (b) $V=0$ for the on-site energy $-5\leq U/t\leq5$ at half filling and $\xi=0$. Note that the well-defined peaks at $k=0$ denote the SF phase. By increasing $U$ the SF phase disappears which is in agreement with Fig.~1. (c) and (d) are the DMRG pair momentum distributions as a function of $k$ for the same systems as (a) and (b), respectively, but at $\xi=0.5$. The system is a spin polarized phase and well-defined peaks occur at $k_{FFLO}$.
}\label{fig5}
\end{figure}

We investigate the oscillatory character of the pair
correlation function by studying the Fourier transform of
the function. Its Fourier transform is given by
\begin{equation}
 G_{pair}(k)=\frac{1}{2L}\sum_{i,j}C_{ij}e^{ik(i-j)}
\end{equation}
The peak of the pair momentum distribution is an indication of a
long-range order pair correlation in the system for different
$\xi$.

The pair momentum distribution can determine the limitation of the SF (for the case where $\xi=0$)
or FFLO phase in the phase diagram. The pair momentum distribution for different states and different interactions is
illustrated in Fig.~\ref{fig5} at half filling. We obtain a similar structure at quarter filling. For $U<0$ and $V>0$ the pair momentum distribution
has a sharp peak that disappears at a certain value
of the dipole-dipole interaction, $V_c$. The value of $V_c$ increases with
increasing $U$. For $V<0$, the pair momentum distribution is a
constant function in the Fourier space.

The ground-state, for the unpolarized case, is the SF state
characterized by a sharp peak centered at momentum $k=0$ in the
pair momentum distribution $G_{pair}(k)$ (Fig.~\ref{fig5}(a)). For $\xi \neq
0$, the ground-state is a 1D FFLO state with $k \neq 0$ (Fig.~\ref{fig5}(c)). Then the $k$ value can be understood as an order
parameter of the FFLO phase. The momentum of the FFLO state
$k_{FFLO}$, at which the $G_{pair}$ shows a strong peak, is
$k_{FFLO}=\pi N \xi/L$. We notice that, in Fig.~\ref{fig5}(d), the value of
$k_{FFLO}$ remains constant for different interaction strengths but
it increases when the filling of atoms increases to the
half filling value. The pair momentum distribution function for quarter filling behaves like the half filling, however
its tail as a function of the momentum is different.

\begin{figure}
\includegraphics[width=0.9\linewidth]{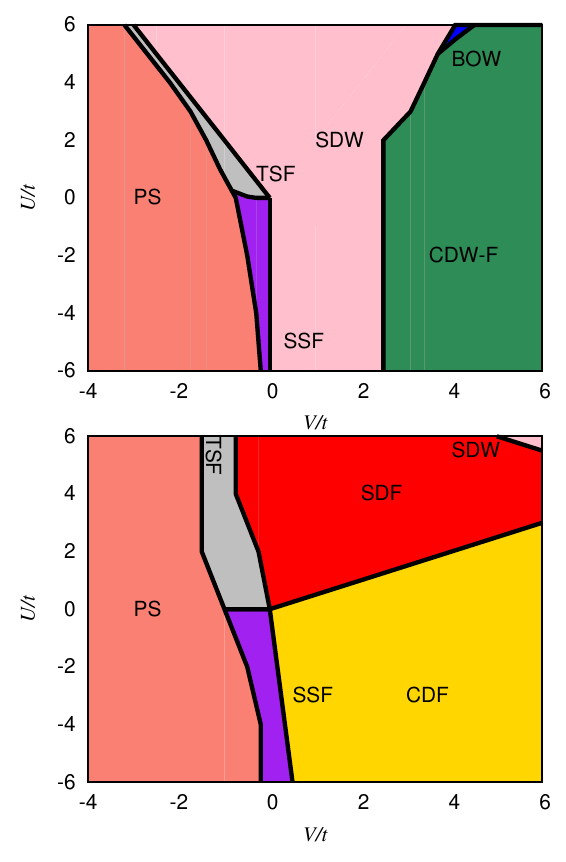}
\caption{(Color online) DMRG phase diagrams of the fermion atoms with dipolar interactions for a $\xi=0.5$ 1D lattice chain. At half filling (top panel), a quantum TSF phase is similar to the case of $\xi=0$, however the spin density wave expands toward negative $U$. The charge density wave is modifies by a mixed state of a combination of the charge density wave and narrow domains of the ferromagnetic state. Moreover, at quarter filling ( bottom panel), rich quantum phases occur and the charge density fluctuation expands with larger $V$ values. The mesh in the horizontal axis is $\Delta V=0.05t$ and the different phases are separated by lines where their thickness is much larger than numerical errors. }
\label{fig6}
\end{figure}

\begin{figure}
\includegraphics[width=0.9\linewidth]{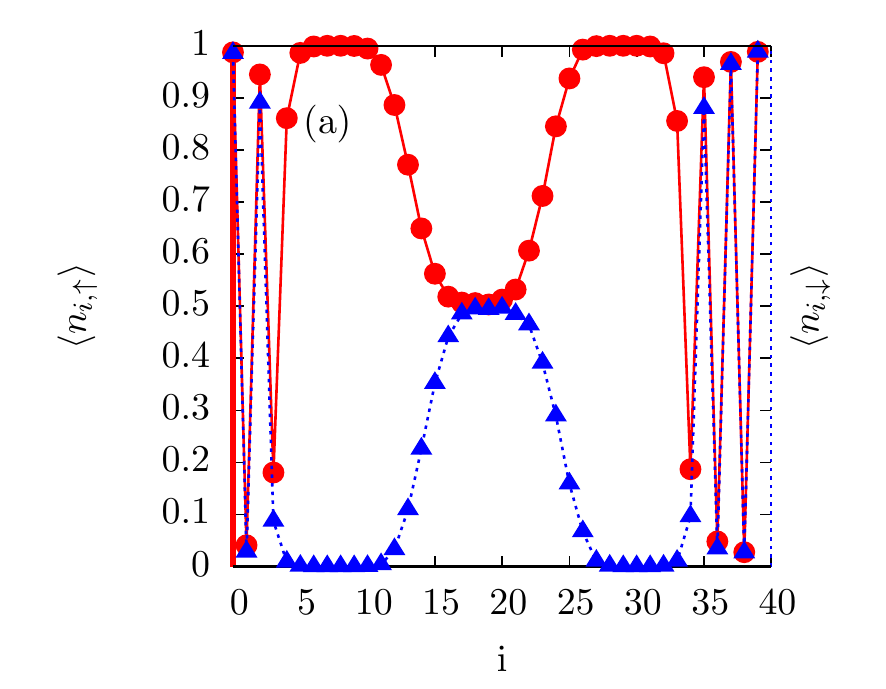}
\includegraphics[width=0.9\linewidth]{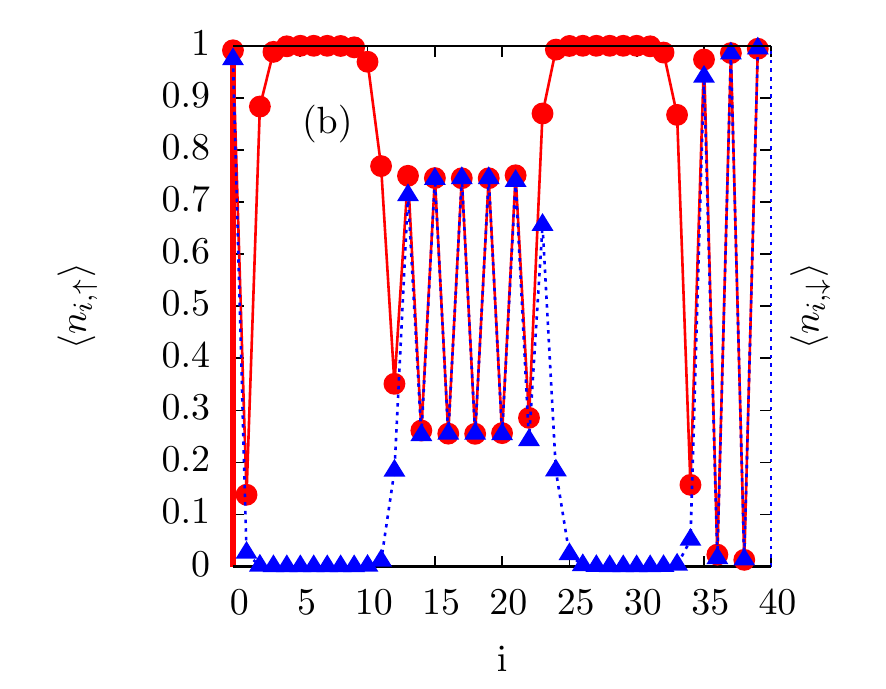}
\caption{(Color online) Density profile of the particles as a function of lattice site for a CDW-F mixed state with $V/t=4$ and (a) $U/t=2$ and (b) $U/t=-2$. A CDW structure for each spin channel in which the spin fluctuation is totally suppressed at a short distance is seen for $U>0$. On the other hand, for $U<0$, the spin fluctuation dose have a noticeable effect at short and close neighborhood but the systems tend to the CDW-F state at larger distances. }
\label{fig7}
\end{figure}

\begin{figure}
\includegraphics[width=0.9\linewidth]{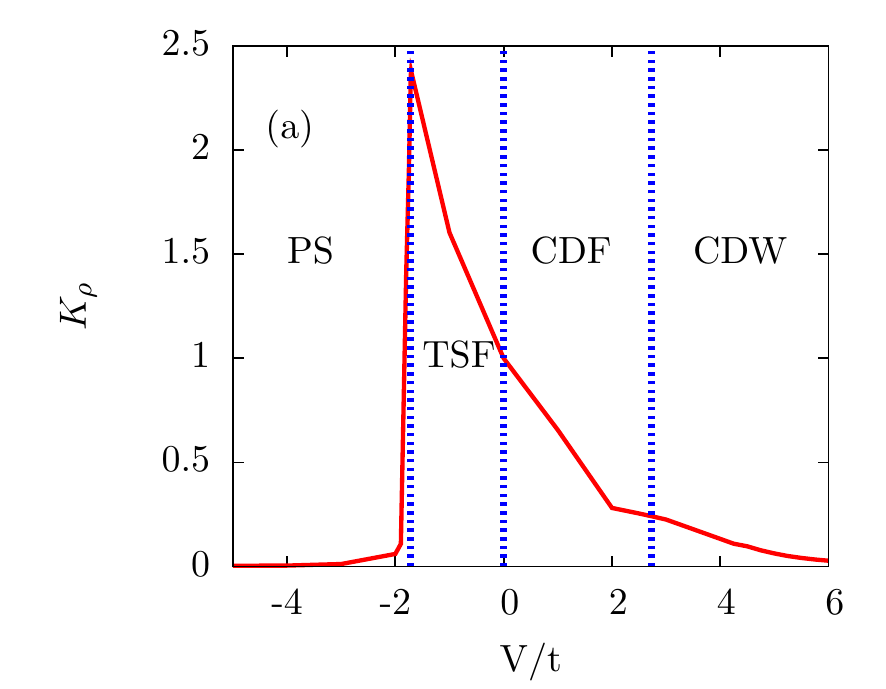}
\includegraphics[width=0.9\linewidth]{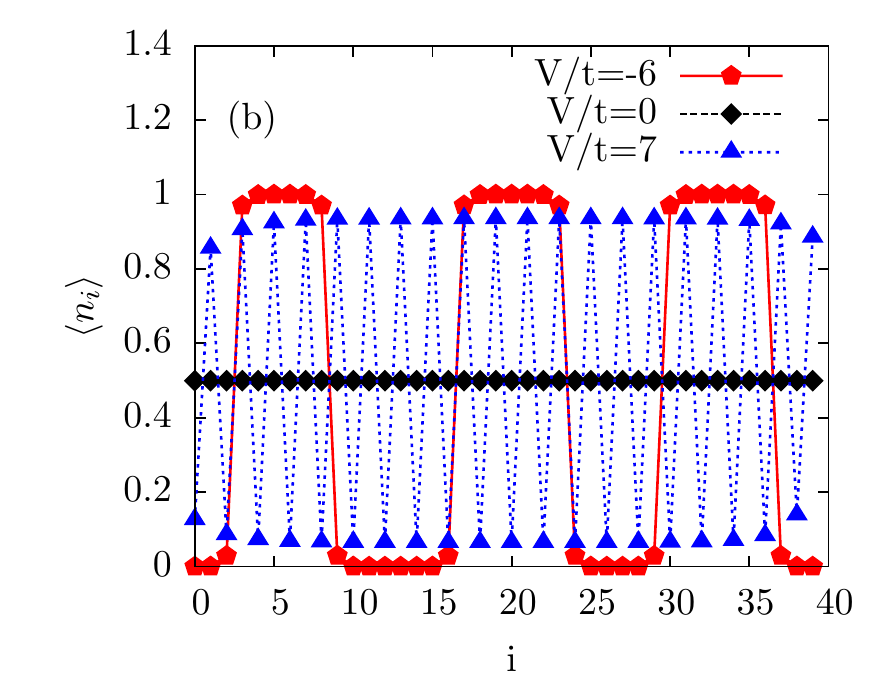}
\caption{(Color online) Luttinger parameter, $K_{\rho}$ as a function of the potential interaction (in units of $t$) at quarter filling for a fully polarized case, $\xi=1$ in the thermodynamic limit. The four stable phases (PS, CDF, CDW and TSF) are shown. Note that the results do not depend on the on-site energy value. (b) A density profile of the particles as a function of lattice site for different phases as shown in (a) at $U=0$.}
\label{fig8}
\end{figure}

The SF or FFLO phase in the fermionic system is the result of
the condensation of the pairs of fermions. If the total spin of the pair
is zero, the state of the two fermions will be a singlet state superfluidity, however the two fermions maybe paired in a triplet state superfluidity. Both the singlet and triple states superfluidity extend to a wide range around $V=0$ due to the presence of the hopping term (see Fig. 1 and Fig. 6 for $\xi=0.5$). For the case of the repulsive interaction, the system undergoes a
quantum phase transition from the SF to an insulator state
. This can be understood by noting that in the strongly interacting regime, density
fluctuations become energetically costly and are therefore
suppressed.

We examine the phase diagram at finite $\xi=0.5$ ($N_{\uparrow}>N_{\downarrow}$) and the results are illustrated in Fig.~\ref{fig6}. The quantum TSF phase is similar to the case of $\xi=0$ at half filling, however the spin density wave expands toward the negative $U$. A small bond order  wave region has taken place for larger $U$ and $V$ values. The charge density wave is modified by a mixed state of a combination of the charge density wave and narrow domains of a ferromagnetic state (CDW-F), since it is energetically favorable.

In order to explore this phase, we show the density profiles of the half filled system for two spin channels at $\xi=0.5$ in Fig.~\ref{fig7} for (a) $U/t=2$ and (b) $U/t=-2$ corresponding to CDW-F illustrated in Fig.~\ref{fig6}. There are two major features in these results. First, there is a CDW structure for each spin channel in which the spin fluctuation is totally suppressed at a short distance, and second, there is a phase shift in their oscillatory structures in such a way that in $U>0$, the maximum of $<n_{i,\uparrow}>$ lies on the minimum of the $<n_{i,\downarrow}>$ and vice versa. Therefore, it shows a mixed state of the CDW phase together with the ferromagnetic state. However, for $U<0$, the spin fluctuation dose have a noticeable effect at short and close neighborhoods but the system tends to the CDW-F state at larger distances. At quarter filling, on the other hand, there are large amounts of quantum fluctuations and the CDF expands with the larger values of $V$. These features originate from a competition between the on-site and the interaction energies for imbalanced particle densities.

Finally, the system in the half filling phase and in the fully polarized case is a ferromagnetic Heisenberg chain model, however there is a rich diagram phase at quarter filling as shown in Fig.~\ref{fig8}. The four stable phases (PS, CDF, CDW and TSF) are obtained. To determine the phase diagram in this case, we calculate the $K_{\rho}$ (=$K_{\sigma}$), the charge and spin gaps given by Eqs.~(2)-(7) in the thermodynamic limit. The value of $K_{\rho}$ is mostly smaller than unity for the range of the value of $V>0$ shown in Fig.~\ref{fig8}(a). The CDF phase for which the hopping energy is dominated, is obtained by conditions in which $\Delta_s\neq 0$ however $\Delta_c=0$. The PS phase has taken place for the large attractive interaction energy with different hole concentrations. The density profile of the different phases is shown in Fig.~\ref{fig8}(b). The particle-hole density profile at the large attractive interaction potential is a constant value $n=0.5$ at $V=0$ and then emerges to the charge density wave at the large repulsive interaction potential.

\section{Conclusions}\label{sec:summary}
We have determined with quite good accuracy the ground-state phase diagrams of the fermion atoms with on-site and dipolar interactions in a 1D lattice at half and quarter filling within the extended Hubbard model by utilizing DMRG approach and finite size scaling.

The model presents a rich phase diagram, depicted in Figs.~1 and 6, illustrating most relevant quantum phases in the $(U,V)$ plane in the 1D lattice of the dipolar system. The competition between the on-site energy, the dipole-dipole interaction and the hopping energy and besides their quantum fluctuations generates different exotic phases in the system. We have elaborated a paring phase in a large area of the repulsive interaction potential at quarter filling.

We have shown, at a given spin polarization, the existence of six phases in the phase diagram and found that they are sensitive to the spin degrees of freedom. In the half filling state, we have found the charge and spin density waves, phase separation, and triplet and single superfluid phases in an unpolarized system, however the charge density wave phase changes qualitatively when the spin degree of freedom is $1/2$ and other phases only change quantitatively in the $(U,V)$ plane. In the quarter filling case, on the other hand, we have found the spin density wave, phase separation, triplet and singlet superfluid and charge and spin density fluctuations phases at finite $\xi$ values, whereas we have only found four stable phases, (phase separation, superfluid state, charge density fluctuation and charge density wave) in the fully spin polarized case. Our obtained phase diagrams should be verified by experiments.

\section{Acknowledgments} Useful discussions with S. Ejima, S. Abedinpour and M. R. Bakhtiari are acknowledged. DMRG results were checked by
using the ALPS DMRG application~\cite{code}.


\begin{thebibliography}{30}

\bibitem{Baranov}
M. A. Baranov, M. Dalmonte, G. Pupillo, and P. Zoller, Condensed Matter Theory of Dipolar Quantum Gases, Chemical Reviews {\bf 112}, 5012 (2012); D.S. Jin and J. Ye, Polar molecules in the quantum regime, Physics Today {\bf 64}, 27 (2011); T. Lahaye1, C. Menotti, L. Santos, M. Lewenstein and T. Pfau, The physics of dipolar bosonic quantum gases, 
Rep. Prog. Phys. {\bf 72}, 126401 (2009); M. A. Baranov, Theoretical progress in many-body physics with ultracold dipolar gases, Phys. Report {\bf 464}, 71 (2008).

\bibitem{Leggett1}
J. G. Bednorz, K. A. Mueller, Possible High Tc superconductivity in the Ba-La-Cu-O system, Z. Phy. B {\bf 64}, 189 (1986); A. L. Leggett, Nat. Phys. What DO we know about high Tc, {\bf 2}, 134 (2006).

\bibitem{Leggett}
A. L. Leggett, Bose-Einstein condensation in the alkali gases: Some fundamental concepts, Rev. Mod. Phys. {\bf 73}, 307 (2001).

\bibitem{Imada}
Masatoshi Imada, Atsushi Fujimori, and Yoshinori Tokura, Metal-insulator transitions, Rev. Mod. Phys. {\bf 70}, 1039 (1998).
\bibitem{fflo}
P. Flude and R. A. Ferrel, Superconductivity in a Strong Spin-Exchange Field, Phys. Rev, {\bf 135}, A550 (1964);
A. J. Larkin and Y. N. Ovchinikov, Nonuniform state of superconductors, Zh. Eksp. Teor. {\bf 47} 1136
(1964).
\bibitem{Cowley}
R. A. Cowley, Structural phase transitions I. Landau theory, Adv. in Phys. {\bf 29}, 1, (1980).

\bibitem{Sansone}
C. Bruder, Rosario Fazio, and Gerd Sch\"{o}n, Superconductor–Mott-insulator transition in Bose systems with finite-range interactions, Phys. Rev. B {\bf 47}, 342 (1993); B. Capogrosso-Sansone, C. Trefzger, M. Lewenstein, P. Zoller, and G. Pupillo, Quantum Phases of Cold Polar Molecules in 2D Optical Lattices, Phys. Rev. Lett. {\bf 104}, 125301 (2010); K. G\'{o}ral, L. Santos, and M. Lewenstein, Quantum Phases of Dipolar Bosons in Optical Lattices {\it ibid} {\bf 88}, 170406 (2002).


\bibitem{Kumar}
M. Kumar, S. Sarkar, S. Ramasesha, Quantum Phases of Long Range 1-D Bose-Hubbard Model: Field Theoretic and DMRG Study at Different Densities arXiv:0812.5059; A G Volosniev, J R Armstrong, D V Fedorov, A S Jensen, M Valiente and N T Zinner, Bound states of dipolar bosons in one-dimensional systems, New J. Phys. {\bf 15} 043046 (2013).
\bibitem{1D-gas}
Xi-Wen Guan, Murray T. Batchelor, and Chaohong Lee, Fermi gases in one dimension: From Bethe ansatz to experiments, Rev. Mod. Phys. {\bf 85}, 1633 (2013);  Yean-an Liao, Ann Sophie C. Rittner, Tobias Paprotta, Wenhui Li, Guthrie B. Partridge, Randall G. Hulet, Stefan K. Baur	
and Erich J. Mueller, Spin-imbalance in a one-dimensional Fermi gas, Nature (London), {\bf 467}, 567 (2010); Tomasz Sowin'ski, Omjyoti Dutta, Philipp Hauke, Luca Tagliacozzo, and Maciej Lewenstein
Phys. Rev. Lett. {\bf 108}, 115301 (2012); T. St{\"{o}}ferle, H. Moritz, K. G{\"{u}}nter, M. K{\"{o}}hl, and T. Esslinger, Dipolar Molecules in Optical Lattices, Phys. Rev. Lett. {\bf 96}, 030401 (2006).

\bibitem{huang}
Yi-Ping Huang, Daw-Wei Wang, Quantum phase diagrams of fermionic dipolar gases in a planar array of one-dimensional tube,  Phys. Rev. A {\bf 80}, 053610 (2009).
\bibitem{Uchino}
S. Uchino, A. Tokuno and T. Giamarchi, Unconventional superfluidity in quasi-one-dimensional systems, Phys. Rev. A {\bf 89}, 023623 (2014).
\bibitem{Exp}
T. Lahaye, T. Koch, B. Fr\"{o}hlich, M. Fattori, J. Metz, A. Griesmaier, S. Giovanazzi and T. Pfau, Strong dipolar effects in a quantum ferrofluid, Nature (London) {\bf 448}, 672 (2007)

\bibitem{Exp0}
K.-K. Ni, S. Ospelkaus, M. H. G. de Miranda, A. P\'{e}er, B. Neyenhuis, J. J. Zirbel, S. Kotochigova, P. S. Julienne, D. S. Jin, J. Ye, A High Phase-Space-Density Gas of Polar Molecules, Science, {\bf 322}, 231 (2008); J. Stuhler, A. Griesmaier, T. Koch, M. Fattori, T. Pfau, S. Giovanazzi, P. Pedri, and L. Santos, Observation of Dipole-Dipole Interaction in a Degenerate Quantum Gas, Phys. Rev. Lett. {\bf 95}, 150406 (2005).

\bibitem{Exp1}
 Bo Yan, Steven A. Moses, Bryce Gadway, Jacob P. Covey, Kaden R. A. Hazzard, Ana Maria Rey, Deborah S. Jin and Jun Ye, Observation of dipolar spin-exchange interactions with lattice-confined polar molecules, Nature (London) {\bf 501}, 521 (2013); Mingwu Lu, Nathaniel Q. Burdick, and Benjamin L. Lev, Quantum Degenerate Dipolar Fermi Gas, Phys. Rev. Lett. {\bf 108}, 215301 (2012).
 \bibitem{Exp2}
 Amodsen Chotia, Brian Neyenhuis, Steven A. Moses, Bo Yan, Jacob P. Covey, Michael Foss-Feig, Ana Maria Rey, Deborah S. Jin, and Jun Ye, Long-Lived Dipolar Molecules and Feshbach Molecules in a 3D Optical Lattice, Phys. Rev. Lett. {\bf 108}, 080405 (2012).
\bibitem{code}
B Bauer et al., The ALPS project release 2.0: open source software for strongly correlated systems, J. Stat. Mech. 2011, P05001 (2011).
\bibitem{Torre}
E. G. Dalla Torre, E. Berg, and E. Altman, Hidden Order in 1D Bose Insulators, Phys. Rev. Lett. {\bf 97}, 260401 (2006); S. R. White, Density matrix formulation for quantum renormalization groups, {\it ibid}, {\bf 69}, 2863 (1992); U. Schollwoeck, The density-matrix renormalization group, Rev. Mod. Phys. {\bf 77}, 259 (2005).
\bibitem{Hubbard}
J. Hubbard, Electron Correlations in Narrow Energy Bands, Proc. R. Soc. Lond. A {\bf 276}, 238 (1963).
\bibitem{EHM}
N. Bartolo, D. J. Papoular, L. Barbiero, C. Menotti, and A. Recati, Dipolar-induced resonance for ultracold bosons in a quasi-one-dimensional optical lattice,
Phys. Rev. A {\bf 88}, 023603 (2013);  K. Sano and Y. \={O}no, Charge gap in the one-dimensional extended Hubbard model at quarter filling, 
Phys. Rev. B {\bf 75}, 113103 (2007).
\bibitem{Nakamura}
M. Nakamura, Tricritical behavior in the extended Hubbard chains, Phys. Rev. B {\bf 61}, 16377 (2000); P. Senguta, A. W. Sandvik, D. K. Campell, Bond-order-wave phase and quantum phase transitions in the one-dimensional extended Hubbard, Phys. Rev. B {\bf 65}, 155113 (2002).
\bibitem{trefzger}
C. Trefzger, C. Menotti, M. Lewenstein, Pair-Supersolid Phase in a Bilayer System of Dipolar Lattice Bosons, Phys. Rev. Lett. {\bf 103}, 035304 (2009).
\bibitem{asgari}
S. Abedinpour, R. Asgari and M. Polini, Theory of correlations in strongly interacting fluids of two-dimensional dipolar bosons, Phys. Rev. A {\bf 86}, 043601 (2012); S. Abedinpour, R. Asgari, B. Tanatar and M. Polini, Ground-state and dynamical properties of two-dimensional dipolar Fermi liquids, Ann. Phys. {\bf 340}, 25 (2014).
\bibitem{boundary}
T. D. K\"{u}hner, S. R. White, H. Monien, One-dimensional Bose-Hubbard model with nearest-neighbor interaction, Phys. REv. B {\bf 61}, 12474 (2000).
\bibitem{Giamarchi}
T. Giamarchi, {\it Quantum Physics in One Dimendion} (Clarendon,
Oxford, 2004).
\bibitem{Clay}
R. Torsten Clay, Anders W. Sandvik, and David K. Campbell, Possible exotic phases in the one-dimensional extended Hubbard model, Phys. Rev. B {\bf 59}, 4665 (1999).
\bibitem{schulz1}
H. J. Schulz, Correlation exponents and the metal-insulator transition in the one-dimensional Hubbard model, Phys. Rev. Lett. {\bf 64}, 2831 (1990); S. Ejima, F. Gebhard, S. Nishimoto and Y. Ohta, Phase diagram of the t-U-V$_1$-V$_2$ model at quarter filling,  Phys. Rev. B {\bf 72}, 033101 (2005).
\bibitem{penc}
K. Penc and F. Mila, Phase diagram of the one-dimensional extended Hubbard model with attractive and/or repulsive interactions at quarter filling, Phys. Rev. B {\bf 49}, 9670 (1994).
\bibitem{zoller}
M. Dalmonte, G. Pupillo and P. Zoller, One-Dimensional Quantum Liquids with Power-Law Interactions: The Luttinger Staircase, Phys. Rev. Lett. {\bf 105}, 140401 (2010).

\bibitem{schulz}
K. S. Bedell, Z. Wang, D. E. Meltzer, {\it Strongly correlated Electronic Materials}, (Addision-Wesley, New Yourk, 1994); H. J. Schulz, arXiv:9412036.
\bibitem{Sogo}
T. Sogo, L. He, T. Miyakawa, S. Yi, H. Lu and H. Pu, Dynamical properties of dipolar Fermi gases, New J. Phys. {\bf 11}, 055017 (2008); S. Ronen, J. L. Bohn, Zero sound in dipolar Fermi gases, Phys. Rev. A {\bf 81}, 033601 (2010).

\bibitem{white}
S. R. White, I. Affleck, and D. J. Scalapino, Friedel oscillations and charge density waves in chains and ladders, Phys. Rev. B {\bf 65}, 165122 (2002); S. Ejima and S. Nishimoto, Phase Diagram of the One-Dimensional Half-Filled Extended Hubbard Model, Phys. Rev. Lett. {\bf 99}, 216403 (2007).
\bibitem{Silva}
T. N. De Silva, Phase diagram of two-component dipolar fermions in one-dimensional optical lattices, Phys. Lett. A {\bf 377}, 871 (2013).

\bibitem{tsuchiizu}
M. Tsuchiizu and A. Furusaki, Phase Diagram of the One-Dimensional Extended Hubbard Model at Half Filling, Phys. Rev. Lett. {\bf 88}, 056402 (2002); M. M\'{e}nard and C. Bourbonnais, Renormalization group analysis of the one-dimensional extended Hubbard model, Phys. Rev. B {\bf 83}, 075111 (2011); S. Ejima and S. Nishimoto, Ground-state properties of the one-dimensional extended Hubbard model at half filling,
J. Phys. Chem. Solids {\bf 69}, 3293 (2008).
\bibitem{all}
R. Casalbuoni and G. Nardulli, Inhomogeneous superconductivity in condensed matter and QCD, Rev. Mod. Phys. {\bf 76}, 263
(2004); Martin W. Zwierlein, André Schirotzek, Christian H. Schunck, Wolfgang Ketterle, Fermionic Superfluidity with Imbalanced Spin Populations, Science {\bf 311}, 492 (2006); G. B. Partridge, W. Li,     R. I. Kamar, Y.-a Liao, R. G. Hulet, Pairing and Phase Separation in a Polarized Fermi Gas {\it ibid}, {\bf 311}, 503 (2006); Martin W. Zwierlein, Christian H. Schunck, André Schirotzek and Wolfgang Ketterle, Direct observation of the superfluid phase transition in ultracold Fermi gases, Nature (London), {\bf 442}, 54, (2006); M. M. Parish, F. M. Marchetti, A. Lamacraft and B. D. Simons, Finite-temperature phase diagram of a polarized Fermi condensate, Nat. Phys. {\bf 3}, 124 (2007).

\end{thebibliography}
\end{document}